\documentclass[journal=jctcce,manuscript=article]{achemso}
\usepackage[version=3]{mhchem}

\usepackage{amssymb,amsmath,amsfonts,multicol,multirow,longtable,array}
\usepackage{lscape}
 
\usepackage{appendix}
\usepackage{soul} 
\usepackage[usenames]{color}
\usepackage{makecell}
\usepackage{enumerate}
\usepackage{xr}
\usepackage[T1]{fontenc} 
\usepackage{booktabs}
\usepackage{tikz}
\usepackage{threeparttable}
\usepackage{graphicx}
\usepackage[figuresright]{rotating}
\usepackage{amsfonts}
\usepackage{colortbl}
\usepackage{framed}
\usepackage{verbatim}
\usepackage{amsbsy}
\usepackage{bm}
\usepackage{bbm}
\usepackage[normalem]{ulem}
\usepackage{float}
\usepackage[linesnumbered,boxed]{algorithm2e}
\usepackage{algorithm2e}
\usepackage{diagbox}
\usepackage{tabularx} 
\usepackage{booktabs} 
\usepackage{physics}
\usepackage{subcaption}

\usepackage[colorlinks=true, urlcolor=blue]{hyperref}
\usepackage{enumitem}
\setlist[enumerate]{label=(\arabic*), itemsep=2pt, parsep=0pt, topsep=3pt, partopsep=0pt}
\usepackage{xcolor}

\usepackage{braket}

\newcounter{xformulation}

\newfloat{Algorithm}{htbp}{alg}
\floatname{Algorithm}{Algorithm}

\newcounter{exe}[figure]
\newcommand{\iexe}{\refstepcounter{exe}\the\value{exe}:}

\setkeys{acs}{maxauthors = 0} 

\author{Xiaoyu Zhang} 
\email{zhangxiaoyu@stu.pku.edu.cn} 
\affiliation{College of Chemistry and Molecular Engineering, Peking University, Beijing 100871, the People's Republic of China}

\title{A Unified Formulation for $\langle \hat{S}^2 \rangle $ in Two-Component TDDFT}

\begin{document}

\begin{abstract}
Two-component linear-response time-dependent density functional theory (TDDFT) provides a unified framework that encompasses noncollinear excitations in noncollinear reference states, as well as both spin-conserving and spin-flip excitations in collinear reference states. In this work, we present a general formalism for evaluating the expectation value $\langle \hat{S}^2 \rangle$ of electronically excited states obtained within two-component TDDFT. We then derive and analyze specialized forms of the resulting equations for collinear reference determinants, for which the two-component formalism decomposes into conventional spin-conserving and spin-flip TDDFT. The resulting working equations are systematically compared with previously proposed theoretical approaches. On the basis of our analysis, $\langle \hat{S}^2 \rangle$ in the excited states is shown to arise from two distinct sources: (i) $\langle \hat{S}^2 \rangle_0$ in the reference state and (ii) additional $\Delta\langle \hat{S}^2 \rangle$ introduced by the excitation process itself. Finally, we evaluate the expectation value $\langle \hat{S}^2 \rangle$ by performing two-component TDDFT calculations based on two-component DFT, unrestricted Kohn–Sham (UKS), and restricted open-shell Kohn–Sham (ROKS) reference states, respectively.
\end{abstract}

\section{Introduction}
Linear-response TDDFT is prevalent due to its balance between efficiency and accuracy.\cite{tddft} In the non-relativistic limit, two-component TDDFT is an adequate framework even in a noncollinear magnetic system \cite{RN69}. Also, two-component TDDFT is a suitable framework for describing relativistic electrons. \cite{10.1063/1.1436462,10.1063/1.2137315} On a collinear reference state, non-relativistic two-component TDDFT is divided into spin-conserving and spin-flip TDDFT.   
Spin-flip TDDFT, formulated on an open-shell reference state, has been demonstrated to be particularly useful for the description of conical intersections, charge-transfer excited states, nonadiabatic derivative couplings, spin–orbit couplings, and bond dissociation processes.\cite{doi:10.1021/acs.jctc.5c01272} 

In order to further enable applications in spectroscopy, it is vital to know the spin multiplicity of excited states; that is, to calculate $\langle \hat{S}^2 \rangle$. The operators $\hat{S}^2$ and $\hat{S}_z$ commute with the spin-free exact electronic Hamiltonian; therefore, both correspond to conserved observables in these cases. In contrast, for a spin-dependent Hamiltonian, $\hat{S}^2$ and $\hat{S}_z$ are always not conserved observables. In the nonrelativistic limit or scalar relativity, owing to the intrinsically single-determinantal character of Kohn–Sham density functional theory (KS-DFT), solutions of two-component DFT are, in general, not eigenfunctions of either $\hat{S}^2$ or $\hat{S}_z$, \cite{doi:10.1021/acs.jctc.7b00832} whereas unrestricted Kohn–Sham (UKS) solutions are typically not eigenfunctions of $\hat{S}^2$. Solutions of two-component TDDFT sometimes have spin contamination due to the incompleteness of spin space. Thus, calculating $\langle \hat{S}^2 \rangle$ is important for analyzing spin multiplicities and spin contamination in the non-relativistic limit and scalar relativity. When we consider full special relativity, $\langle \hat{S}^2 \rangle$ is still useful. In the context of nonadiabatic molecular dynamics, Zhou and Shu utilize the expectation value $\langle \hat{S}^2 \rangle$ as a quantitative descriptor of spin mixing, whereby the degree of singlet–triplet admixture is positively correlated with the associated electronic hopping probability. \cite{doi:10.1021/acs.jpca.4c04639}
However, calculating $\langle \hat{S}^2 \rangle$ is not trivial and hasn't been systematically explored in two-component TDDFT. 

Maurice and Head-Gordon \cite{https://doi.org/10.1002/qua.560560840} derived analytical expressions for the expectation value of the total spin-squared operator, $\langle \hat{S}^2 \rangle$, within the framework of spin-conserving configuration interaction singles (CIS). Ipatov \textit{et al.} \cite{IPATOV200960} subsequently obtained expressions for $\langle \hat{S}^2 \rangle$ in spin-conserving TDDFT. Myneni and Casida \cite{MYNENI201772} addressed and resolved several ambiguities associated with the evaluation of $\langle \hat{S}^2 \rangle$ in spin-conserving TDDFT and proposed an \textit{ab initio} formulation. Building on the conceptual framework introduced by Ipatov \textit{et al.}, Li \textit{et al.} \cite{10.1063/1.3573374} derived the corresponding working equations for spin-flip TDDFT, which they reported in the appendix of their work.

To enrich and systematize the theoretical framework in this area, we introduce a unified formulation for the spin-squared expectation value, $\langle \hat{S}^2 \rangle$, within two-component time-dependent density functional theory (TDDFT). We begin from the second-quantized representation of $\hat{S}^2$ and demonstrate that its algebraic structure closely parallels that of the second-quantized electronic Hamiltonian $\hat{H}$. On this basis, we derive explicit expressions for $\langle \hat{S}^2 \rangle$ evaluated over general Slater determinants, which are found to be analogous to the corresponding energy expectation values derived from the Slater–Condon rules.
Subsequently, we formulate two-component TDDFT in the second-quantized equation-of-motion (EOM) framework, which is formally equivalent to the Casida equation in the first-quantized matrix representation. By substituting $\hat{H}$ with $\hat{S}^2$ in the EOM-like TDDFT expressions, we obtain a Casida-type eigenvalue equation in the first-quantized representation for the evaluation of spin-squared expectation values. In close analogy to the electronic excitation energies $\omega$ ($\Delta \langle \hat{H} \rangle$), the changes in spin-squared expectation values, $\Delta \langle \hat{S}^2 \rangle$, can be computed within this Casida-like formalism.
We then specialize the treatment to collinear reference states and derive further simplified working equations, which enable direct comparison with previously proposed spin-analysis schemes. Finally, we present numerical results for a range of molecular systems to illustrate and benchmark the proposed methodology.

\section{Theory}
\subsection{Second-Quantized $\hat{S}^2$}
We use  $\Gamma, \Lambda, \Theta...$ for two-component AO basis functions, $P, Q, R...$ for unspecified two-component molecular orbitals, and  $I, J, K...$ / $A,B,C...$ for occupied / virtual orbitals. We use $g_\mu,g_\nu,...$ to denote the spatial part of AO basis functions. The corresponding spin-up basis is denoted as:
\begin{equation}
    |\mu \rangle = \begin{pmatrix}
g_\mu\\
0
    \end{pmatrix}
\end{equation}
And the spin-down basis is denoted as:
\begin{equation}
    |\bar{\mu} \rangle = \begin{pmatrix}
0\\
g_\mu
    \end{pmatrix}
\end{equation}
Also, we use the corresponding lowercase letters to denote the spin-up or spin-down parts of two-component orbitals:
\begin{equation}
    |P\rangle = \begin{pmatrix}
        p\\
        \bar{p}
    \end{pmatrix}
    =\sum_\mu
    \begin{pmatrix}
        c_{\mu p} g_\mu\\
        c_{\bar{\mu}\bar{p}} g_\mu
    \end{pmatrix}
\end{equation}
We denote the coefficient matrix as follows:
\begin{equation}
 (  C^{\alpha})_{\mu p} = c_{\mu p}
\end{equation}
\begin{equation}
  ( C^{\beta})_{\mu p} = c_{\bar{\mu} \bar{p}}
\end{equation}
We denote the column $p$ of $C^\alpha$ as $C^\alpha_p$, and that of $C^\beta$ as $C^\beta_p$.

We use short notations for double-electron integrals:
\begin{equation}
    (PQ|RS) = \int d \mathbf{r}_1 d \mathbf{r}_2 \frac{P^\dagger(\mathbf{r}_1) Q(\mathbf{r}_1)R^\dagger(\mathbf{r}_2) S(\mathbf{r}_2)}{|\mathbf{r}_1-\mathbf{r}_2|} 
\end{equation}
\begin{equation}
    \langle PQ|RS \rangle= (PR|QS)
\end{equation}
\begin{equation}
    \langle PQ| |RS \rangle =   \langle PQ |RS \rangle -   \langle PQ| SR \rangle
\end{equation}

We denote the matrix representation of a one-electron spin operator  $\hat{s}_u$ as:
\begin{equation}
    (s_u)_{PQ} = \langle P| \frac{1}{2} \hat{\sigma}_u |Q \rangle
\end{equation}
Here, $u=x,y,z$. $\hat{\sigma}_u$ is the Pauli matrix along the $u$-axis.

Because $\hat{s}_u$ is a one-body operator, its second-quantized form is:
\begin{equation}
    \hat{s}_u = \sum_{PQ} (s_u)_{PQ} \hat{a}_P^\dagger \hat{a}_Q
\end{equation}
The total spin of the electrons is defined as:
\begin{equation}
    \hat{S}^2 = \sum_u \hat{s}_u \hat{s}_u 
\end{equation}
By substituting the $\hat{s}_u$, we obtain:
\begin{equation}
       \hat{S}^2 = \sum_u  \sum_{PQRS} (s_u)_{PQ} (s_u)_{RS} \hat{a}^\dagger_P \hat{a}_Q \hat{a}_R^\dagger \hat{a}_S
\end{equation}
There are two basic anti-commutation relations:
\begin{equation}
    \{ \hat{a}_P, \hat{a}_Q \} = 0
\end{equation}
\begin{equation}
    \{ \hat{a}_P, \hat{a}_Q^\dagger \} = \delta_{PQ}
\end{equation}
Using the two relations, we obtain:
\begin{equation}
    \hat{S}^2 = \sum_u  \sum_{PQRS} (s_u)_{PQ} (s_u)_{RS}  (\delta_{QR} \hat{a}_P^\dagger \hat{a}_S+ \hat{a}_P^\dagger \hat{a}_R^\dagger \hat{a}_S \hat{a}_Q)
\end{equation}
We denote:
\begin{equation}
    W_{PQRS}=2 \sum_u (s_u)_{PR} (s_u)_{QS}
\end{equation}
Based on the definition, we have:
\begin{equation}
    W_{PQRS}=W_{QPSR} = W^*_{RSPQ}=W^*_{SRQP}
\end{equation}
Finally, we obtain:
\begin{equation}
    \hat{S}^2 = \frac{3}{4} \sum_P \hat{a}_P^\dagger \hat{a}_P + \frac{1}{2} \sum_{PQRS} W_{PQRS} \hat{a}_P^\dagger \hat{a}_Q^\dagger \hat{a}_S \hat{a}_R
\end{equation}

\subsection{Expectation Values of $\hat{S}^2$}
We denote $|X\rangle$ as the reference single slater determinant wavefunction $|\Psi\rangle_0$, constructed by a set of orthonormal molecular orbitals. We usually obtain $|\Psi\rangle_0$ using the two-component Hartree-Fock method. A singly excited wavefunction is denoted as $|Y \rangle=\hat{a}^\dagger_A \hat{a}_I |X \rangle$. A doubly excited wavefunction is denoted as $|Z \rangle = \hat{a}^\dagger_A \hat{a}_B^\dagger \hat{a}_J \hat{a}_I |X \rangle$. Based on the Slater-Condon rule \cite{PhysRev.34.1293,PhysRev.36.1121}, we can calculate the expectation value of $\hat{S}^2$ for different determinants.
\begin{equation}
    \langle X| \hat{S}^2 |X \rangle = \frac{3}{4}N +\frac{1}{2} \sum_{IJ} (W_{IJIJ}-W_{IJJI})
    \label{eq:sodef}
\end{equation}
\begin{equation}
      \langle X| \hat{S}^2 |Y \rangle =\sum_J(W_{IJAJ}-W_{IJJA})= \sum_J(W_{AJIJ}-W_{AJJI})^*
      \label{eq:s1def}
\end{equation}
\begin{equation}
      \langle X| \hat{S}^2 |Z \rangle=W_{IJAB}-W_{IJBA} = (W_{ABIJ}-W_{ABJI})^*
      \label{eq:s2def}
\end{equation}
Here, $N$ is the number of electrons. We now need to further rewrite $W_{PQRS}$ with two-component orbital indices to $W_{pqrs}$ with one-component orbital indices. We denote the overlap matrix as:
\begin{subequations}
\begin{equation}
        S_{pq}=\langle p|q \rangle
\end{equation}
\begin{equation}
        S_{\bar{p}q}=\langle \bar{p}|q \rangle
\end{equation}
\begin{equation}
        S_{p\bar{q}}=\langle p|\bar{q} \rangle
\end{equation}
\begin{equation}
        S_{\bar{p} \bar{q}}=\langle \bar{p}|\bar{q} \rangle
\end{equation}
\end{subequations}

By expanding a molecular orbital into a set of basis functions, we have:
\begin{subequations}
    \begin{equation}
    S_{pq} = \sum_{\mu \nu} c^*_{\mu p} c_{\nu q} \langle g_\mu | g_\nu  \rangle = \sum_{\mu \nu} c^*_{\mu p} c_{\nu q} S_{\mu \nu}
\end{equation}
\begin{equation}
    S_{\bar{p}q} = \sum_{\mu \nu} c^*_{\bar{\mu} \bar{p}} c_{\nu q} \langle g_\mu | g_\nu  \rangle = \sum_{\mu \nu} c^*_{\bar{\mu} \bar{p}} c_{\nu q} S_{\mu \nu}
\end{equation}
 \begin{equation}
    S_{p\bar{q}} = \sum_{\mu \nu} c^*_{\mu p} c_{\bar{\nu} \bar{q}} \langle g_\mu | g_\nu  \rangle = \sum_{\mu \nu} c^*_{\mu p} c_{\bar{\nu} \bar{q}} S_{\mu \nu}
\end{equation}
 \begin{equation}
    S_{\bar{p}\bar{q}} = \sum_{\mu \nu} c^*_{\bar{\mu} \bar{p}} c_{\bar{\nu} \bar{q}} \langle g_\mu | g_\nu  \rangle = \sum_{\mu \nu} c^*_{\bar{\mu} \bar{p}} c_{\bar{\nu} \bar{q}} S_{\mu \nu}
\end{equation}
\label{eq:smat}
\end{subequations}

Then,
\begin{equation}
    W_{pqrs} = \frac{1}{2}(S_{pr}-S_{\bar{p}\bar{r}})(S_{qs}-S_{\bar{q}\bar{s}}) + S_{p \bar{r}} S_{\bar{q}s} +S_{\bar{p}r} S_{q\bar{s}}
    \label{eq:wdef}
\end{equation}
By introducing eq. \ref{eq:wdef} into eq. \ref{eq:sodef}, we produce the same result as eq. (19) in Ref. \citenum{ss_2cghf}. But we need to derive eq. \ref{eq:sodef} under a common AO representation for more convenient implementation. We define the first-order reduced density matrix as:
\begin{equation}
    D_{\Gamma \Lambda} = \sum_I c^*_{\Gamma I} c_{\Lambda I}
\end{equation}
We denote four blocks of the density matrix as:
\begin{subequations}
    \begin{equation}
    D_{\mu \nu} = \sum_i c^*_{\mu i} c_{\nu i} = (D^{\alpha \alpha})_{\mu \nu}
\end{equation}
\begin{equation}
    D_{\bar{\mu} \nu} = \sum_i c^*_{\bar{\mu} \bar{i}} c_{\nu i} = (D^{\beta \alpha})_{\mu \nu}
\end{equation}
\begin{equation}
    D_{\mu \bar{\nu}} = \sum_i c^*_{\mu i} c_{\bar{\nu} \bar{i}} =  (D^{\alpha \beta})_{\mu \nu}
\end{equation}
\begin{equation}
    D_{\bar{\mu} \bar{\nu}} = \sum_i c^*_{\bar{\mu} \bar{i}} c_{\bar{\nu} \bar{i}} =(D^{\beta \beta})_{\mu \nu}
\end{equation}
\label{eq:dmat}
\end{subequations}

After the mathematical derivation shown in Appendix A, we have the expectation value of the total electronic spin in the reference state:
\begin{equation}
\begin{split}
        \langle \hat{S}^2 \rangle_0 &= \frac{3}{4}N +\frac{1}{4}[\mathrm{Tr}(S(D^{\alpha \alpha}-D^{\beta \beta}))]^2 + \mathrm{Tr}(S D^{\alpha \beta})\times \mathrm{Tr}(S D^{\beta \alpha})\\
        &-\frac{1}{4}[\mathrm{Tr}(D^{\alpha \alpha}S D^{\alpha \alpha}S)+\mathrm{Tr}(D^{\beta \beta}S D^{\beta \beta}S)-2\mathrm{Tr}(D^{\alpha \beta}S D^{\beta \alpha}S)] - \mathrm{Tr}(D^{\alpha \alpha}S D^{\beta \beta}S)
        \end{split}
        \label{eq:s0_raw}
\end{equation}
If there are no fractional occupations, we have:
\begin{equation}
    \langle \hat{S}^2 \rangle_0 = \frac{1}{2}N +\frac{1}{4}[\mathrm{Tr}(S(D^{\alpha \alpha}-D^{\beta \beta}))]^2 + \mathrm{Tr}(S D^{\alpha \beta})\times \mathrm{Tr}(S D^{\beta \alpha})+ \mathrm{Tr}(D^{\alpha \beta} S D^{\beta \alpha}S)-\mathrm{Tr}(D^{\alpha \alpha}S D^{\beta \beta} S)
    \label{eq:s0}
\end{equation}
Another way to calculate $\langle \hat{S}^2 \rangle_0$ is to decompose $\hat{S}^2$ into one-body operators $\hat{S}_+$, $\hat{S}_-$, and $\hat{S}_z$, and we have:
\begin{equation}
      \hat{S}^2 = \frac{1}{2}(\hat{S}_+\hat{S}_-+\hat{S}_-\hat{S}_+) +\hat{S}_z^2
\end{equation}
\begin{equation}
    \begin{split}
        \langle\hat{S}_+\hat{S}_- \rangle_0= \mathrm{Tr}(D^{\alpha \alpha}S) +\mathrm{Tr}(D^{\beta \alpha}S)\mathrm{Tr}(D^{\alpha \beta}S)-\mathrm{Tr}[(D^{\alpha \alpha}S)(D^{\beta \beta}S)]
\end{split}
\end{equation}
\begin{equation}
        \langle\hat{S}_-\hat{S}_+ \rangle_0= \mathrm{\mathrm{Tr}}(D^{\beta \beta}S) +\mathrm{\mathrm{Tr}}(D^{\beta \alpha}S)\mathrm{\mathrm{Tr}}(D^{\alpha \beta}S)-\mathrm{\mathrm{Tr}}[(D^{\alpha \alpha}S)(D^{\beta \beta}S)]
\end{equation}
\begin{equation}
\begin{split}
        \langle\hat{S}_z\hat{S}_z \rangle_0  = \frac{1}{4}\mathrm{Tr}(D^{\alpha \alpha}S)+\frac{1}{4}\mathrm{Tr}(D^{\beta \beta}S) + \frac{1}{4} [\mathrm{Tr}(D^{\alpha \alpha }S)-\mathrm{Tr}(D^{\beta \beta}S)]^2 \\- \frac{1}{4} \mathrm{Tr}(D^{\alpha \alpha}SD^{\alpha \alpha}S)- \frac{1}{4} \mathrm{Tr}(D^{\beta \beta}SD^{\beta \beta}S) +\frac{1}{2}\mathrm{Tr}(D^{\alpha \beta}SD^{\beta \alpha}S)
\end{split}
\end{equation}
This method produces the same result as eq. \ref{eq:s0_raw}.
Here, $D$ and $S$ are both under a common AO representation.

After the mathematical derivation shown in Appendix B, we have:
\begin{equation}
\begin{split}
     \langle X| \hat{S}^2 | Y \rangle &=  [\left(C^\alpha_a \right)^\dagger
     \left( \frac{1}{2}S \mathrm{Tr}\left( S \left(D^{\alpha \alpha}-D^{\beta \beta} \right) \right)- \frac{1}{2}\left( S \left(D^{\alpha \alpha} \right)^T S \right) -\left( S \left(D^{\beta \beta} \right)^T S \right) \right) \left(C^\alpha_i \right) \\
     &+\left( C_a^\alpha \right)^\dagger      \left(S \mathrm{Tr}\left( S D^{\beta \alpha} \right)  + \frac{1}{2} S \left(D^{\beta \alpha} \right)^T S\right) \left( C_i^\beta \right)\\
       &+\left( C_a^\beta \right)^\dagger      \left(S \mathrm{Tr}\left( S D^{ \alpha \beta} \right)  + \frac{1}{2} S \left(D^{\alpha \beta } \right)^T S\right) \left( C_i^\alpha \right)\\
       &+ \left(C^\beta_a \right)^\dagger
     \left( -\frac{1}{2}S \mathrm{Tr}\left( S \left(D^{\alpha \alpha}-D^{\beta \beta} \right) \right)- \frac{1}{2}\left( S \left(D^{\beta \beta} \right)^T S \right) -\left( S \left(D^{\alpha \alpha} \right)^T S \right) \right) \left(C^\beta_i \right)]^*
\end{split}
\label{eq:s1}
\end{equation}
For eq. \ref{eq:s2def}, we directly use the definition of $W$ in eq. \ref{eq:wdef}, and we have:
\begin{equation}
\begin{split}
       \langle X| \hat{S}^2 |Z \rangle &= \{\frac{1}{2} [(S_{ai}-S_{\bar{a}\bar{i}})(S_{bj}-S_{\bar{b}\bar{j}})- (S_{aj}-S_{\bar{a}\bar{j}})(S_{bi}-S_{\bar{b}\bar{i}})]\\
       &+S_{a\bar{i}} S_{\bar{b}j} +S_{\bar{a}i} S_{b \bar{j}} - S_{a \bar{j}} S_{\bar{b}i} - S_{\bar{a}j} S_{b \bar{i}}\}^*
\end{split}
\end{equation}
This equation can be easily implemented by using eq. \ref{eq:smat}.

\subsection{$\Delta \langle \hat{S}^2 \rangle$ of Two-Component TDDFT in Second-Quantized EOM}

The formalism of the equation-of-motion (EOM) focuses on a transition operator $O_\lambda^\dagger$ \cite{RevModPhys.40.153}, which changes an eigenstate $|0\rangle$ of the Hamiltonian to another eigenstate $|\lambda \rangle$:
\begin{equation}
    O_\lambda^\dagger |0\rangle =|\lambda \rangle
\end{equation}
Also, it satisfies the killer condition:
\begin{equation}
    O_\lambda |0 \rangle = 0
\end{equation}
$O_\lambda^\dagger$ is determined by one of the following three conditions:
\begin{equation}
    \langle 0| \delta O_\lambda [\hat{H},O_\lambda^\dagger]|0\rangle = \omega_\lambda   \langle 0| \delta O_\lambda O_\lambda^\dagger|0\rangle 
\label{eq:eom1}
\end{equation}
\begin{equation}
    \langle 0| [\delta O_\lambda, [\hat{H},O_\lambda^\dagger]]|0\rangle = \omega_\lambda   \langle 0| [\delta O_\lambda ,O_\lambda^\dagger]|0\rangle 
    \label{eq:eom2}
\end{equation}
\begin{equation}
    \langle 0| [\delta O_\lambda, \hat{H},O_\lambda^\dagger]|0\rangle = \omega_\lambda   \langle 0| [\delta O_\lambda, O_\lambda^\dagger]|0\rangle 
    \label{eq:eom3}
\end{equation}
Here, $\omega_\lambda$ is the excitation energy. It is noteworthy that the three conditions are equivalent only if $|0 \rangle$ is exact, the killer condition is satisfied, and $O_\lambda$ is expanded on a complete basis. The triple commutator is defined as:
\begin{equation}
    [A,B,C] = \frac{1}{2} ([[A,B],C]+[A,[B,C]])
\end{equation}
RPA in this work uses the second EOM condition, which is eq. \ref{eq:eom2}.
We express the two-component RPA in second-quantized EOM:

\begin{equation}
    M X = \omega N X
    \label{eq:eomw}
\end{equation}
\begin{equation}
    M_{PQRS}= \langle [ O_{PQ}, [\hat{H},O_{RS}^\dagger]] \rangle_0
\end{equation}
\begin{equation}
    N_{PQRS}= \langle [ O_{PQ},O_{RS}^\dagger] \rangle_0
\end{equation}
\begin{equation}
    O^\dagger_{PQ} = \hat{a}_P^\dagger \hat{a}_Q    
\end{equation}
Here, $\hat{a}^\dagger$ and $\hat{a}$ are the creation and annihilation operators. By expanding $\hat{O}$ on a set of single-excitation operators and single-deactivation operators, we have two cases to discuss: $PQRS=AIBJ$ and $PQRS=AIJB$. Based on the Slater-Condon rule, we have:
\begin{equation}
    M_{AIBJ} = f_{AB} \delta_{IJ} - f_{JI} \delta_{AB} +K_{AIBJ}
\end{equation}
\begin{equation}
    M_{AIJB}= -K_{AIJB}
\end{equation}
\begin{equation}
    N_{AIBJ} = \delta_{AB} \delta_{IJ}
\end{equation}
\begin{equation}
    N_{AIJB}=0
\end{equation}
\begin{equation}
    f_{PQ} = h_{PQ} + \sum_{R \in \Phi_0} \bra{PR}  \ket{QR}
\end{equation}
\begin{equation}
    K_{PQRS}=\bra{PS} \ket{QR}
\end{equation}
We can easily prove two symmetries:
\begin{equation}
    M_{PQRS} = M_{QPSR}^*
\end{equation}
\begin{equation}
    N_{PQRS} =- N_{QPSR}^*
\end{equation}
We can present the Casida equation in the two-component RPA:
\begin{equation}
    \begin{pmatrix}
        M_{AIBJ} & M_{AIJB}\\
        M_{AIJB}^* &   M_{AIBJ}^*
    \end{pmatrix}
    X= \omega
    \begin{pmatrix}
        1 & 0\\
        0 & -1
    \end{pmatrix}
    X 
\end{equation}
We rewrite the Fock and kernel as:
\begin{equation}
     f_{PQ}= \frac{\partial E}{\partial D_{PQ}}
\end{equation}
\begin{equation}
        K_{PQRS}= \frac{\partial^2 E}{\partial D_{PQ}\partial D_{SR}}
\end{equation}
By translating $E$ from HF energy to DFT energy, we can translate two-component RPA to two-component TDDFT. In two-component TDDFT, Fock and kernel can be expressed as:
\begin{equation}
f_{PQ}= h_{PQ} + \sum_{R \in \Phi_0}(\langle PR|QR \rangle -c_{\mathrm{HF}}  \langle PR|RQ \rangle ) + \frac{\partial E_{\mathrm{xc}}}{\partial D_{PQ}}
\end{equation}
\begin{equation}
    K_{PQRS} = \langle PS |QR \rangle - c_{\mathrm{HF}} \langle PS | RQ \rangle +  \frac{\partial^2 E_{\mathrm{xc}}}{\partial D_{PQ}\partial D_{SR}}
\end{equation}

The excitation energy $\omega$ can be seen as $\Delta \langle \hat{H} \rangle$. Motivated by this, we can also calculate $\Delta \langle \hat{S}^2 \rangle$ via similar equations.
\begin{equation}
    M^s X^s = \Delta \langle\hat{S}^2 \rangle N^s X^s
    \label{eq:eoms}
\end{equation}
\begin{equation}
    M^s_{PQRS}= \langle [ O_{PQ}, [\hat{S}^2,O_{RS}^\dagger]] \rangle_0
\end{equation}
\begin{equation}
    N^s_{PQRS}= \langle [ O_{PQ},O_{RS}^\dagger] \rangle_0
\end{equation}
Obviously, 
\begin{equation}
    N^s_{PQRS}= N_{PQRS}
\end{equation}
The two symmetries are still preserved:
\begin{equation}
    M^s_{PQRS} = (M_{QPSR}^s)^*
    \label{eq:symmMs}
\end{equation}
\begin{equation}
    N^s_{PQRS} =- (N_{QPSR}^s)^*
\end{equation}
We denote the 'Fock' and 'kernel' for $\hat{S}^2$ as:
\begin{equation}
    f_{PQ}^s =\sum_{R \in \Phi_0} W_{PRQR} -W_{PRRQ}
\end{equation}
\begin{equation}
    K_{PQRS}^s = W_{PSQR}-W_{PSRQ }
\end{equation}
Based on the Slater-Condon rule, we have:
\begin{equation}
    M^s_{AIBJ} = f^s_{AB} \delta_{IJ} - f^s_{JI} \delta_{AB} +K^s_{AIBJ}
    \label{eq:msadef}
\end{equation}
\begin{equation}
    M^s_{AIJB}= -K^s_{AIJB}
    \label{eq:msdadef}
\end{equation}
Following the similar idea of deriving eq. \ref{eq:s1}, we have:
\begin{equation}
\begin{split}
       f^s_{PQ}&=  [\left(C^\alpha_q \right)^\dagger
     \left( \frac{1}{2}S \mathrm{Tr}\left( S \left(D^{\alpha \alpha}-D^{\beta \beta} \right) \right)- \frac{1}{2}\left( S \left(D^{\alpha \alpha} \right)^T S \right) -\left( S \left(D^{\beta \beta} \right)^T S \right) \right) \left(C^\alpha_p \right) \\
     &+\left( C_q^\alpha \right)^\dagger      \left(S \mathrm{Tr}\left( S D^{\beta \alpha} \right)  + \frac{1}{2} S \left(D^{\beta \alpha} \right)^T S\right) \left( C_p^\beta \right)\\
       &+\left( C_q^\beta \right)^\dagger      \left(S \mathrm{Tr}\left( S D^{ \alpha \beta} \right)  + \frac{1}{2} S \left(D^{\alpha \beta } \right)^T S\right) \left( C_p^\alpha \right)\\
       &+ \left(C^\beta_q \right)^\dagger
     \left( -\frac{1}{2}S \mathrm{Tr}\left( S \left(D^{\alpha \alpha}-D^{\beta \beta} \right) \right)- \frac{1}{2}\left( S \left(D^{\beta \beta} \right)^T S \right) -\left( S \left(D^{\alpha \alpha} \right)^T S \right) \right) \left(C^\beta_p \right)]^*
\end{split}
\label{eq:focks}
\end{equation}
By directly introducing the definition of $W$ in eq. \ref{eq:wdef}, we have:
\begin{equation}
\begin{split}
     K_{PQRS}^s &= \frac{1}{2} [(S_{pq}-S_{\bar{p}\bar{q}})(S_{sr}-S_{\bar{s}\bar{r}})- (S_{pr}-S_{\bar{p}\bar{r}})(S_{sq}-S_{\bar{s}\bar{q}})]\\
       &+S_{p\bar{q}} S_{\bar{s}r} +S_{\bar{p}q} S_{s \bar{r}} - S_{p \bar{r}} S_{\bar{s}q} - S_{\bar{p}r} S_{s \bar{q}}
\end{split}
\label{eq:kernels}
\end{equation}
We can use eq. \ref{eq:eoms} to calculate $\Delta\langle \hat{S}^2 \rangle$ of the eigenvectors in eq. \ref{eq:eomw}.
\begin{equation}
    \Delta\langle \hat{S}^2 \rangle = \frac{X^\dagger M^s X}{X^\dagger N X}
\end{equation}
$X$ is one of the eigenvectors solved from eq. \ref{eq:eomw}. The theory will be more complete if the two equations eq. \ref{eq:eoms} and eq. \ref{eq:eomw} share the same set of eigenvectors. Unfortunately, this result will be true only when the operator basis is complete. Proof is shown in Appendix C.

Electronic states exhibiting vanishing excitation energies require special consideration. Zero-eigenvalue modes correspond to orbital rotations that are not associated with any restoring force. These modes can be partitioned into two distinct categories: (i) proper modes, which originate from degeneracies of the underlying wave functions and whose symmetry restoration does not necessarily lead to an energy lowering, and (ii) improper modes, which arise from symmetry breaking and for which symmetry restoration always results in a reduction of the total energy. \cite{10.1063/1.4824905} Proper modes can be chosen to give \(X^\dagger N X > 0\), but improper modes can only give  \(X^\dagger N X = 0\). \cite{COLPA1986377} Such improper modes induce numerical instabilities in the computed values of \(\Delta \langle \hat{S}^2 \rangle\). Because these improper modes are physically meaningless, we enforce the condition \(\Delta \langle \hat{S}^2 \rangle = 0\) for these improper modes and suggest that they don't need to be analyzed.

Eq. \ref{eq:eomw} has pairwise solutions. And physical solutions are preserved based on:
\begin{equation}
    X^\dagger N X >0
\end{equation}
If we rewrite $X$ to be $\begin{pmatrix}
    \tilde{X}\\
    \tilde{Y}
\end{pmatrix}$, this inequality becomes:
\begin{equation}
    |\tilde{X}| > |\tilde{Y}|
\end{equation}
\subsection{$\langle \hat{S}^2 \rangle$ on a Collinear Reference }
A collinear reference is usually acquired by UKS or ROKS. By rewriting spin-up and spin-down molecular orbitals from UKS or ROKS in the form of two-component molecular orbitals, we still preserve the orthonormality of the molecular orbitals. Thus, the Slater-Condon rule still works, and we can largely use the aforementioned equations.

Starting from eq. \ref{eq:s0_raw}, we have the expectation value of the total electron spin in a collinear reference state:
\begin{equation}
    \begin{split}
        \langle \hat{S}^2 \rangle_0 &= \frac{3}{4}N +\frac{1}{4}[\mathrm{Tr}(S(D^{\alpha \alpha}-D^{\beta \beta}))]^2 \\
        &-\frac{1}{4}[\mathrm{Tr}(D^{\alpha \alpha}S D^{\alpha \alpha}S)+\mathrm{Tr}(D^{\beta \beta}S D^{\beta \beta}S)] - \mathrm{Tr}(D^{\alpha \alpha}S D^{\beta \beta}S)
        \end{split}
\end{equation}
If there are no fractional occupations, we have:
\begin{equation}
      \langle \hat{S}^2 \rangle_0 = \frac{1}{2}N +\frac{1}{4}[\mathrm{Tr}(S(D^{\alpha \alpha}-D^{\beta \beta}))]^2 -\mathrm{Tr}(D^{\alpha \alpha}S D^{\beta \beta} S)
\end{equation}
This equation reproduces the result in eq. (143) of ref. \citenum{PhysRev.97.1490}.

Starting from eq. \ref{eq:focks}, we have a simplified expression for 'Fock' of spin:
\begin{equation}
\begin{split}
       f^s_{PQ}&=  [\left(C^\alpha_q \right)^\dagger
     \left( \frac{1}{2}S \mathrm{Tr}\left( S \left(D^{\alpha \alpha}-D^{\beta \beta} \right) \right)- \frac{1}{2}\left( S \left(D^{\alpha \alpha} \right)^T S \right) -\left( S \left(D^{\beta \beta} \right)^T S \right) \right) \left(C^\alpha_p \right) \\
       &+ \left(C^\beta_q \right)^\dagger
     \left( -\frac{1}{2}S \mathrm{Tr}\left( S \left(D^{\alpha \alpha}-D^{\beta \beta} \right) \right)- \frac{1}{2}\left( S \left(D^{\beta \beta} \right)^T S \right) -\left( S \left(D^{\alpha \alpha} \right)^T S \right) \right) \left(C^\beta_p \right)]^*
\end{split}
\end{equation}
Equivalently, it can be rewritten as:
\begin{equation}
\begin{split}
     f_{PQ}^s &= \frac{1}{2} S_{pq} \sum_j (S_{jj}-S_{\bar{j}\bar{j}}) -\frac{1}{2} S_{\bar{p}\bar{q}} \sum_j (S_{jj}-S_{\bar{j}\bar{j}})\\
&-\frac{1}{2}\sum_j S_{pj}S_{jq} -\sum_j S_{p\bar{j}} S_{\bar{j}q} -\frac{1}{2} \sum_j S_{\bar{p}\bar{j}} S_{\bar{j}\bar{q}} -\sum_j S_{\bar{p}j} S_{j\bar{q}}
\end{split}
\end{equation}
For a two-component orbital $P$, we use $P_\alpha$ to denote it if it only has the spin-up part; and $P_\beta$ to denote it if it only has the spin-down part. Thus, the 'Fock' of spin on a collinear reference can be further simplified.
\begin{equation}
\begin{split}
     f^s_{P_\alpha Q_\alpha} &= \frac{1}{2} S_{pq} \sum_j (S_{jj}-S_{\bar{j}\bar{j}}) -\frac{1}{2} \sum_j S_{pj}S_{jq} -\sum_j S_{p\bar{j}}S_{\bar{j}q}  \\
     &= \frac{1}{2} \delta_{pq}\sum_j (S_{jj}-S_{\bar{j}\bar{j}}) - \frac{1}{2} n_p\delta_{pq}-\sum_j S_{p\bar{j}}S_{\bar{j}q}
\end{split}
\end{equation}
\begin{equation}
    f^s_{P_\alpha Q_\beta} =  f^s_{P_\beta Q_\alpha} = 0 
\end{equation}
\begin{equation}
\begin{split}
     f^s_{P_\beta Q_\beta} &= -\frac{1}{2} S_{\bar{p} \bar{q}} \sum_j (S_{jj}-S_{\bar{j}\bar{j}}) -\frac{1}{2} \sum_j S_{\bar{p}\bar{j}}S_{\bar{j}\bar{q}} -\sum_j S_{\bar{p}j}S_{j\bar{q}}\\
     &= -\frac{1}{2} \delta_{pq} \sum_j (S_{jj}-S_{\bar{j}\bar{j}}) -\frac{1}{2}n_{\bar{p}} \delta_{pq}-\sum_j S_{\bar{p}j}S_{j\bar{q}}
\end{split}
\end{equation}
Starting from eq. \ref{eq:kernels}, we have simplified expressions for the 'kernel' of spin:
\begin{equation}
K^s_{P_\alpha Q_\alpha R_\alpha S_\alpha}
= \frac{1}{2}\Big( S_{pq}\,S_{sr}-S_{pr}\,S_{sq}\Big) = \frac{1}{2}(\delta_{pq} \delta_{sr}- \delta_{pr} \delta_{sq})
\end{equation}

\begin{equation}
K^s_{P_\beta Q_\beta R_\beta S_\beta}
= \frac{1}{2}\Big( S_{\bar{p}\bar{q}}\,S_{\bar{s}\bar{r}}-S_{\bar{p}\bar{r}}\,S_{\bar{s}\bar{q}}\Big) = \frac{1}{2}(\delta_{pq} \delta_{sr}- \delta_{pr} \delta_{sq})
\end{equation}

\begin{equation}
K^s_{P_\alpha Q_\alpha R_\beta S_\beta}
= -\frac{1}{2}\,S_{pq}\,S_{\bar{s}\bar{r}} \;-\; S_{p\bar{r}}\,S_{\bar{s}q} = 
-\frac{1}{2}\delta_{pq} \delta_{sr}- S_{p\bar{r}}\,S_{\bar{s}q}
\end{equation}

\begin{equation}
K^s_{P_\beta Q_\beta R_\alpha S_\alpha}
= -\frac{1}{2}\,S_{\bar{p}\bar{q}}\,S_{sr} \;-\; S_{\bar{p}r}\,S_{s\bar{q}}
=-\frac{1}{2}\delta_{pq} \delta_{sr} - S_{\bar{p}r}\,S_{s\bar{q}}
\end{equation}

\begin{equation}
K^s_{P_\alpha Q_\beta R_\alpha S_\beta}
= \frac{1}{2}\,S_{pr}\,S_{\bar{s}\bar{q}} \;+\; S_{p\bar{q}}\,S_{\bar{s}r}
=\frac{1}{2}\delta_{pr} \delta_{sq} + S_{p\bar{q}}\,S_{\bar{s}r}
\end{equation}

\begin{equation}
K^s_{P_\beta Q_\alpha R_\beta S_\alpha}
= \frac{1}{2}\,S_{\bar{p}\bar{r}}\,S_{sq} \;+\; S_{\bar{p}q}\,S_{s\bar{r}}
=\frac{1}{2}\delta_{pr} \delta_{sq} + S_{\bar{p}q}\,S_{s\bar{r}}
\end{equation}
The left ten terms are zero.

By using eq. \ref{eq:msadef} and eq. \ref{eq:msdadef}, we can simplify the $M^s$:
\begin{equation}
M^s=
\begingroup\footnotesize
\begin{pmatrix}
M^s_{A_\alpha I_\alpha B_\alpha J_\alpha} & M^s_{A_\alpha I_\alpha B_\beta J_\beta} & M^s_{A_\alpha I_\alpha J_\alpha B_\alpha} & M^s_{A_\alpha I_\alpha J_\beta B_\beta} & 0 & 0 & 0 & 0\\
M^s_{A_\beta I_\beta B_\alpha J_\alpha} & M^s_{A_\beta I_\beta B_\beta J_\beta} & M^s_{A_\beta I_\beta J_\alpha B_\alpha} & M^s_{A_\beta I_\beta J_\beta B_\beta} & 0 & 0 & 0 & 0\\
M^s_{I_\alpha A_\alpha B_\alpha J_\alpha} & M^s_{I_\alpha A_\alpha B_\beta J_\beta} & M^s_{I_\alpha A_\alpha J_\alpha B_\alpha} & M^s_{I_\alpha A_\alpha J_\beta B_\beta} & 0 & 0 & 0 & 0\\
M^s_{I_\beta A_\beta B_\alpha J_\alpha} & M^s_{I_\beta A_\beta B_\beta J_\beta} & M^s_{I_\beta A_\beta J_\alpha B_\alpha} & M^s_{I_\beta A_\beta J_\beta B_\beta} & 0 & 0 & 0 & 0\\
0 & 0 & 0 & 0 & M^s_{A_\alpha I_\beta B_\alpha J_\beta} & M^s_{A_\alpha I_\beta J_\alpha B_\beta} & 0 & 0\\
0 & 0 & 0 & 0 & M^s_{I_\alpha A_\beta B_\alpha J_\beta} & M^s_{I_\alpha A_\beta J_\alpha B_\beta} & 0 & 0\\
0 & 0 & 0 & 0 & 0 & 0 & M^s_{A_\beta I_\alpha B_\beta J_\alpha} & M^s_{A_\beta I_\alpha J_\beta B_\alpha}\\
0 & 0 & 0 & 0 & 0 & 0 & M^s_{I_\beta A_\alpha B_\beta J_\alpha} & M^s_{I_\beta A_\alpha J_\beta B_\alpha}
\end{pmatrix}
\endgroup
\end{equation}
The 16 left-upper blocks are spin-conserving transitions. The 16 right-lower blocks are spin-flip transitions. The $4\times4$ spin-flip block is block-diagonal with 8 non-zero submatrices. Detailed expressions for spin-conserving blocks are as follows:

\begin{equation}
    M^s_{A_\alpha I_\alpha B_\alpha J_\alpha} = \delta_{ab} (\sum_k S^*_{i \bar{k}}S_{j \bar{k}})- \delta_{ij} (\sum_k S^*_{b \bar{k}}S_{a \bar{k}})
\end{equation}
\begin{equation}
    M^s_{A_\alpha I_\alpha B_\beta J_\beta} = - S^*_{i \bar{j}} S_{a \bar{b}}
\end{equation}
\begin{equation}
    M^s_{A_\beta I_\beta B_\alpha J_\alpha} = -S^*_{b\bar{a}} S_{j \bar{i}}
\end{equation}
\begin{equation}
    M^s_{A_\beta I_\beta B_\beta J_\beta} = \delta_{ab} (\sum_k S^*_{k\bar{j}}S_{k \bar{i}} ) - \delta_{ij} (\sum_k S^*_{k \bar{a}}S_{k \bar{b}}) 
\end{equation}
\begin{equation}
    M^s_{A_\alpha I_\alpha J_\alpha B_\alpha} =0
\end{equation}
\begin{equation}
    M^s_{A_\alpha I_\alpha J_\beta B_\beta} = S_{i \bar{b}}^* S_{a \bar{j}}
\end{equation}
\begin{equation}
    M^s_{A_\beta I_\beta J_\alpha B_\alpha} = S^*_{j\bar{a}} S_{b \bar{i}}
\end{equation}
\begin{equation}
    M^s_{A_\beta I_\beta J_\beta B_\beta}=0
\end{equation}
The other 8 spin-conserving blocks can be acquired by using eq. \ref{eq:symmMs}. 
Also, by virtue of this symmetry, two spin-flip blocks have a simple relationship:
\begin{equation}
\begin{pmatrix}
M^s_{A_\beta I_\alpha\, B_\beta J_\alpha} & M^s_{A_\beta I_\alpha\, J_\beta B_\alpha}\\
M^s_{I_\beta A_\alpha\, B_\beta J_\alpha} & M^s_{I_\beta A_\alpha\, J_\beta B_\alpha}
\end{pmatrix}
=
\begin{pmatrix}
0 & 1\\
1 & 0
\end{pmatrix}
\left[
\begin{pmatrix}
M^s_{A_\alpha I_\beta\, B_\alpha J_\beta} & M^s_{A_\alpha I_\beta\, J_\alpha B_\beta}\\
M^s_{I_\alpha A_\beta\, B_\alpha J_\beta} & M^s_{I_\alpha A_\beta\, J_\alpha B_\beta}
\end{pmatrix}
\right]^{*}
\begin{pmatrix}
0 & 1\\
1 & 0
\end{pmatrix}
\end{equation}
Detailed expressions for spin-flip blocks are as follows:
\begin{equation}
    M^s_{A_\alpha I_\beta\, B_\alpha J_\beta} = \delta_{ij} \delta_{ab} [\sum_k (S_{kk}-S_{\bar{k}\bar{k}})+1] -\delta_{ij} \sum_k S_{a\bar{k}} S_{\bar{k}b} +\delta_{ab} \sum_k S_{\bar{j}k} S_{k\bar{i}} +S_{a\bar{i}}S_{\bar{j}b}
\end{equation}
\begin{equation}
    M^s_{A_\alpha I_\beta\, J_\alpha B_\beta} = -S_{a\bar{i}} S_{\bar{b}j}
\end{equation}
\begin{equation}
    M^s_{I_\alpha A_\beta\, B_\alpha J_\beta} = -S_{i\bar{a}} S_{\bar{j}b}
\end{equation}
\begin{equation}
    M^s_{I_\alpha A_\beta\, J_\alpha B_\beta} = \delta_{ij} \delta_{ab} [-\sum_k (S_{kk}-S_{\bar{k}\bar{k}})+1] -\delta_{ij} \sum_k S_{k\bar{a}} S_{\bar{b}k} +\delta_{ab} \sum_k S_{\bar{k}j} S_{i\bar{k}} +S_{i\bar{a}}S_{\bar{b}j}
\end{equation}

These spin-conserving equations exactly reproduce the results in eqs. (2.42)-(2.45) of ref. \citenum{MYNENI201772}, which are the ab initio and 'best simple' formulas according to the reference. Also, these spin-conserving equations reproduce the results by Ipatov \textit{et al.} \cite{IPATOV200960}. Ref. \citenum{MYNENI201772} stated that, during the course of the initial work, Ipatov once used more \textit{ad hoc} ideas and changed $N^s$ from $\begin{pmatrix}
    1& 0\\
    0 &-1
\end{pmatrix}$ to $\begin{pmatrix}
    1 & 0\\
    0& 1
\end{pmatrix}$, which resulted in the loss of pairwise excitation/deactivation transitions in spin-conserving TDDFT. This transformation is argued by introducing an anticommutator, which was proposed by Lynch \textit{et al.} \cite{LYNCH198269}. Li \textit{et al.} \cite{10.1063/1.3573374} followed similar ideas to Ipatov's formally published ideas and derived equations for spin-flip TDDFT in their appendix. The spin-flip equations we derived are slightly different from Li's. Based on our derivation, eq. (A10) in their work should be:
\begin{equation}
    \Delta\langle\hat{S}^2 \rangle = 2S+\frac{\sum\limits_{ai} \tilde{X}_{a\bar{i}}^* \tilde{X}_{a\bar{i}}+\tilde{Y}_{\bar{a}i}^* \tilde{Y}_{\bar{a}i}}{\sum\limits_{ai} \tilde{X}_{a\bar{i}}^* \tilde{X}_{a\bar{i}}-\tilde{Y}_{\bar{a}i}^* \tilde{Y}_{\bar{a}i}} + \Delta \langle P_{\alpha \beta} \rangle
\end{equation}
And eq. (A12) in their work should be:
\begin{equation}
    \Delta\langle\hat{S}^2 \rangle = -2S+\frac{\sum\limits_{ai} \tilde{X}_{\bar{a}i}^* \tilde{X}_{\bar{a}i}+\tilde{Y}_{a\bar{i}}^* \tilde{Y}_{a\bar{i}}}{\sum\limits_{ai} \tilde{X}_{\bar{a}i}^* \tilde{X}_{\bar{a}i}-\tilde{Y}_{a\bar{i}}^* \tilde{Y}_{a\bar{i}}} + \Delta \langle P_{\alpha \beta} \rangle
\end{equation}

\subsection{Noncollinear Exchange and Correlation Kernels}
We use noncollinear exchange-correlation kernels, the second derivative of exchange-correlation energy functionals, generalized from collinear exchange-correlation functionals by the multicollinear approach. \cite{doi:10.1021/acs.jctc.5c01305,PhysRevResearch.5.013036} We briefly revisit the idea of the multi-collinear approach here. 

The multicollinear approach assumes that noncollinear functionals can be written in the following form:
\begin{equation}
    E_{\mathrm{xc}}[n,\mathbf{m}]=\frac{1}{4\pi}\int_0^{2\pi} \int_0^\pi \left.E_{\mathrm{xc}}^{\mathrm{eff}}[n,s]\right|_{s=\bm{m}\cdot \bm{e}} \sin \theta \,\mathrm{d} \theta \,\mathrm{d}\phi
\end{equation}
where
\begin{equation}
    \bm{e}(\theta,\phi) = (\sin \theta \cos \phi, \sin \theta \sin \phi, \cos \theta)
\end{equation}
It was added that it should have the correct collinear limit:
\begin{equation}
    E_{\mathrm{xc}}[n,(0,0,m_z)] =  E^{\mathrm{col}}_{\mathrm{xc}}[n,m_z]
\end{equation}
we can then have:
\begin{equation}
\int_0^1 E^{\mathrm{eff}}_{\mathrm{xc}}[n,st] dt= E_{\mathrm{xc}}^{\mathrm{col}}[n,s]
\label{eq:mc}
\end{equation}
For eq. \ref{eq:mc}, one solution is:
\begin{equation}
    E^{\mathrm{eff}}_{\mathrm{xc}}[n,s] =  E^{\mathrm{col}}_{\mathrm{xc}}[n,s] + \int \frac{\delta E^{\mathrm{col}}_{\mathrm{xc}}[n,s]}{\delta s(\mathbf{r})}s(\mathbf{r}) d \mathbf{r}
\end{equation}
A principal advantage of employing noncollinear exchange–correlation kernels that are invariant under global spin rotations is that two-component TDDFT yields at least one zero-energy excitation for an open-shell reference state.\cite{doi:10.1021/acs.jctc.5c00714} This property is of fundamental importance. For instance, for a collinear high-spin open-shell reference state $|SS\rangle$, spin-flip TDDFT generates the state $|S(S-1)\rangle$. The corresponding zero-excitation theorem ensures that the degeneracy between $|SS\rangle$ and $|S(S-1)\rangle$ is rigorously preserved.

\section{Results and Discussion}
All calculations were performed using the cc-pVDZ basis set. Reference states were obtained by means of two-component density functional theory (generalized Kohn–Sham, GKS), unrestricted Kohn–Sham (UKS), or restricted open-shell Kohn–Sham (ROKS) \cite{RevModPhys.32.179} formalisms. The linear-response treatment was carried out employing a two-component time-dependent density functional theory (TDDFT) implementation. Excitation energies are reported in Hartree. Geometries of calculated molecules are presented in Appendix D. ROKS calculations were performed using the PySCF program package \cite{10.1063/5.0006074}, whereas all other electronic-structure calculations, including UKS, GKS, and TDDFT, were carried out with our in-house JAX-based quantum chemistry code, IQC.
\begin{table}[H]
    \centering
    \scriptsize
    \setlength{\tabcolsep}{3pt}
    \renewcommand{\arraystretch}{1.1}
    \resizebox{\linewidth}{!}{%
    \begin{tabular}{cc|ccc|ccc|ccc|ccc}
        \hline
        \multicolumn{2}{c|}{} & \multicolumn{3}{c|}{$\mathrm{H_2O(UKS)}$} & \multicolumn{3}{c|}{$\mathrm{H_2O^+(UKS)}$} & \multicolumn{3}{c|}{$\mathrm{H_2O^+(ROKS)}$} & \multicolumn{3}{c}{$\mathrm{H_3(GKS)}$} \\
        \cline{3-14}
        \multicolumn{1}{c}{Functional} & Root
        & $\omega$ & $\Delta\langle\hat S^2\rangle$ & $\langle\hat S^2\rangle$
        & $\omega$ & $\Delta\langle\hat S^2\rangle$ & $\langle\hat S^2\rangle$
        & $\omega$ & $\Delta\langle\hat S^2\rangle$ & $\langle\hat S^2\rangle$
        & $\omega$ & $\Delta\langle\hat S^2\rangle$ & $\langle\hat S^2\rangle$ \\
        \hline

        \multirow{6}{*}{HF}  & 0  & -- & -- & 0.0000 & -- & -- & 0.7561 & -- & -- & 0.7500 & -- & -- & 0.7642 \\
                            & 1  & 0.2999 & 2.0143 & 2.0143 & 0.0870 & 0.0074 & 0.7634 & -0.0182 & 0.0264 & 0.7764 & 0.2853 & 0.5125 & 1.2767 \\
                            & 2  & 0.2999 & 2.0143 & 2.0143 & 0.0910 & 0.0037 & 0.7597 & 0.0796 & 0.0078 & 0.7578 & 0.2853 & 0.5125 & 1.2767 \\
                            & 3  & 0.2999 & 2.0143 & 2.0143 & 0.2429 & 0.0083 & 0.7644 & 0.0806 & 0.0046 & 0.7546 & 0.4366 & 1.7117 & 2.4759 \\
                            & 4  & 0.3368 & 0.0000 & 0.0000 & 0.2555 & 0.0052 & 0.7612 & 0.2387 & 0.0095 & 0.7595 & 0.4387 & 1.7064 & 2.4706 \\
                            & 5  & 0.3738 & 2.0389 & 2.0389 & 0.5041 & 3.0276 & 3.7837 & 0.2419 & 0.0069 & 0.7569 & 0.4671 & 1.5402 & 2.3044 \\
        \hline

        \multirow{6}{*}{SVWN} & 0  & -- & -- & 0.0000 & -- & -- & 0.7517 & -- & -- & 0.7500 & -- & -- & 0.7522 \\
                              & 1  & 0.2499 & 2.0026 & 2.0026 & 0.0745 & 0.0017 & 0.7534 & -0.0041 & 0.0071 & 0.7571 & 0.2765 & 0.5016 & 1.2538 \\
                              & 2  & 0.2499 & 2.0026 & 2.0026 & 0.0766 & 0.0008 & 0.7525 & 0.0715 & 0.0019 & 0.7519 & 0.2765 & 0.5017 & 1.2539 \\
                              & 3  & 0.2499 & 2.0026 & 2.0026 & 0.2215 & 0.0001 & 0.7518 & 0.0750 & 0.0013 & 0.7513 & 0.4605 & 1.7234 & 2.4756 \\
                              & 4  & 0.2725 & 0.0000 & 0.0000 & 0.2224 & 0.0013 & 0.7530 & 0.2193 & 0.0015 & 0.7515 & 0.4610 & 1.7349 & 2.4871 \\
                              & 5  & 0.3233 & 2.0036 & 2.0036 & 0.4559 & 3.0025 & 3.7542 & 0.2200 & 0.0018 & 0.7518 & 0.4767 & 1.5561 & 2.3083 \\
        \hline

        \multirow{6}{*}{PBE} & 0  & -- & -- & 0.0000 & -- & -- & 0.7519 & -- & -- & 0.7500 & -- & -- & 0.7544 \\
                             & 1  & 0.2449 & 2.0046 & 2.0046 & 0.0868 & 0.0034 & 0.7553 & -0.0060 & 0.0079 & 0.7579 & 0.2839 & 0.5025 & 1.2570 \\
                             & 2  & 0.2449 & 2.0046 & 2.0046 & 0.0936 & 0.0019 & 0.7538 & 0.0828 & 0.0034 & 0.7534 & 0.2839 & 0.5026 & 1.2571 \\
                             & 3  & 0.2449 & 2.0046 & 2.0046 & 0.2347 & 0.0031 & 0.7550 & 0.0900 & 0.0024 & 0.7524 & 0.4522 & 1.7273 & 2.4817 \\
                             & 4  & 0.2699 & 0.0000 & 0.0000 & 0.2399 & 0.0014 & 0.7533 & 0.2312 & 0.0033 & 0.7533 & 0.4528 & 1.7235 & 2.4779 \\
                             & 5  & 0.3193 & 2.0070 & 2.0070 & 0.4530 & 3.0051 & 3.7570 & 0.2356 & 0.0027 & 0.7527 & 0.4769 & 1.5238 & 2.2782 \\
        \hline

        \multirow{6}{*}{B3LYP} & 0  & -- & -- & 0.0000 & -- & -- & 0.7522 & -- & -- & 0.7500 & -- & -- & 0.7549 \\
                               & 1  & 0.2533 & 2.0050 & 2.0050 & 0.0799 & 0.0036 & 0.7559 & -0.0067 & 0.0093 & 0.7593 & 0.2820 & 0.5045 & 1.2594 \\
                               & 2  & 0.2533 & 2.0050 & 2.0050 & 0.0887 & 0.0019 & 0.7541 & 0.0762 & 0.0037 & 0.7537 & 0.2820 & 0.5043 & 1.2593 \\
                               & 3  & 0.2533 & 2.0050 & 2.0050 & 0.2293 & 0.0036 & 0.7558 & 0.0850 & 0.0024 & 0.7524 & 0.4539 & 1.7044 & 2.4593 \\
                               & 4  & 0.2799 & 0.0000 & 0.0000 & 0.2372 & 0.0017 & 0.7540 & 0.2266 & 0.0039 & 0.7539 & 0.4557 & 1.7167 & 2.4717 \\
                               & 5  & 0.3294 & 2.0084 & 2.0084 & 0.4642 & 3.0064 & 3.7587 & 0.2325 & 0.0030 & 0.7530 & 0.4755 & 1.5207 & 2.2756 \\
        \hline
    \end{tabular}%
    }
    \caption{Excitation energies $\omega$ (Hartree), spin-squared change $\Delta\langle\hat S^2\rangle$, and $\langle\hat S^2\rangle$ for reported excited roots 1--5 of four systems using different functionals and reference formalisms. Root~0 only provides the reference-state value $\langle\hat S^2\rangle_0$. Excited states with $\omega=0$ are not reported.}
    \label{tab:ssmol}
\end{table}

In Table~\ref{tab:ssmol}, we summarize the two-component linear-response TDDFT results for $\mathrm{H_2O}$, $\mathrm{H_2O^+}$, and $\mathrm{H_3}$ using different reference-state formalisms (GKS, UKS, and ROKS) and exchange–correlation approximations (HF, SVWN, PBE, and B3LYP). For each reference, we report the excitation energies $\omega$ (Hartree) and the corresponding spin information for the lowest five excited roots (roots~1–5), including the spin-squared change $\Delta\langle\hat S^2\rangle \equiv \langle\hat S^2\rangle-\langle\hat S^2\rangle_0$ and the resulting $\langle\hat S^2\rangle$. Root~0 is shown only to provide the reference-state value $\langle\hat S^2\rangle_0$; we report excitation energies only for roots with $\omega\neq 0$. $\mathrm{H_3}$ is a special noncollinear molecule, and its magnetism can be found in figure. (2) of ref. \citenum{doi:10.1021/acs.jctc.5c00714}. The occurrence of a negative excitation root in $\mathrm{H_2O^+}$ when using a ROKS reference requires clarification. For $\mathrm{H_2O^+}$ with a UKS reference, there is in fact a zero-energy excitation with a proper mode, although it was not explicitly listed in the table. This zero excitation corresponds to the spin-flip transition from $|\tfrac{1}{2},\tfrac{1}{2}\rangle$ to $|\tfrac{1}{2},-\tfrac{1}{2}\rangle$. 
Because the ROKS reference is not obtained by diagonalizing the full Fock matrix, this zero-energy excitation is not rigorously maintained. As a consequence, in the ROKS-based calculation the corresponding root is slightly changed and becomes a negative excitation energy. $\mathrm{H_2O}$ and $\mathrm{H_2O^+}$ both have negligible spin contamination, while $\mathrm{H_3}$ has severe spin contamination in its excited states, which should result from the incompleteness of spin space. $\mathrm{H_3}$ has three pairs of zero excitation with improper modes, which are physically trivial.

\section{Conclusion}
In this work, we propose a unified formulation for $\langle\hat{S}^2 \rangle$ in two-component TDDFT. Results are presented in the non-relativistic limit, but the equations are transferable to two-component special relativistic methods, such as X2C. Similar ideas can also be proposed to calculate $\langle\hat{S}^2 \rangle$ or even $\langle\hat{J}^2 \rangle$ in four-component TDDFT. Following the idea of conducting substitutions in second quantization, expectation values of many observables can be calculated within a two-component or even four-component framework.

\begin{acknowledgement}
Xiaoyu Zhang thanks Yunlong Xiao for the valuable discussion.
\end{acknowledgement}

\section*{\textbf{Author Information}}

\textbf{Corresponding Author}
\begin{itemize}[leftmargin=16pt, nosep]
    \item Xiaoyu Zhang - College of Chemistry and Molecular Engineering, Peking University, Beijing, 100871, P. R. China.\\
    
    \url{https://orcid.org/0009-0009-4178-3519}; 
    Email: \href{mailto:zhangxiaoyu@stu.pku.edu.cn}{\texttt{zhangxiaoyu@stu.pku.edu.cn}}
\end{itemize}

\bibliography{main}

\section*{Appendix }
\subsection*{Appendix A: Derivation from Eq. \ref{eq:sodef} to Eq. \ref{eq:s0_raw} and Eq. \ref{eq:s0}}
By the definition of $W$ in eq. \ref{eq:wdef}, we have:

\begin{equation}
    \sum_{ij}W_{ijij} = \frac{1}{2}[\sum_i (S_{ii}-S_{\bar{i}\bar{i}})][\sum_j (S_{jj}-S_{\bar{j}\bar{j}})]+ 2(\sum_i S_{i \bar{i}}) (\sum_j S_{\bar{j}j})
\end{equation}
\begin{equation}
    \sum_{ij}W_{ijji} = \frac{1}{2} (\sum_{ij}S_{ij}S_{ji}+\sum_{ij}S_{\bar{i}\bar{j}}S_{\bar{j}\bar{i}}-2\sum_{ij}S_{ij} S_{\bar{j}\bar{i}}) + 2 \sum_{ij} S_{i \bar{j}} S_{\bar{j}i}
\end{equation}
Using eqs. \ref{eq:smat} and \ref{eq:dmat}, we have:
\begin{equation}
    \sum_{ij}W_{ijij} = \frac{1}{2}[\mathrm{Tr}(S(D^{\alpha \alpha}-D^{\beta \beta}))]^2 +2 \mathrm{Tr}(S D^{\alpha \beta})\times \mathrm{Tr}(S D^{\beta \alpha})
\end{equation}
\begin{equation}
    \sum_{ij} W_{ijji} = \frac{1}{2}[\mathrm{Tr}(D^{\alpha \alpha}S D^{\alpha \alpha} S)+\mathrm{Tr}(D^{\beta \beta}S D^{\beta \beta} S)-2\mathrm{Tr}(D^{\alpha \beta}S D^{\beta \alpha} S) ]+2\mathrm{Tr}(D^{\alpha \alpha}S D^{\beta \beta} S)
\end{equation}

If there are no fractional occupations, we have:
\begin{equation}
    \left[
        \begin{pmatrix}
            D^{\alpha \alpha} & D^{\alpha \beta}\\
            D^{\beta \alpha}  & D^{\beta \beta}
        \end{pmatrix}
        \begin{pmatrix}
            S & 0 \\
            0 & S
        \end{pmatrix}
        \begin{pmatrix}
            D^{\alpha \alpha} & D^{\alpha \beta}\\
            D^{\beta \alpha}  & D^{\beta \beta}
        \end{pmatrix}
    \right]_{\Gamma\Lambda}= \left[
        \begin{pmatrix}
            D^{\alpha \alpha} & D^{\alpha \beta}\\
            D^{\beta \alpha}  & D^{\beta \beta}
        \end{pmatrix}
    \right]_{\Gamma\Lambda}
\end{equation}
then we have:

\begin{equation}
    D^{\alpha \alpha} S D^{\alpha \alpha} +   D^{\alpha \beta} S D^{\beta \alpha} = D^{\alpha \alpha} 
\end{equation}
\begin{equation}
    D^{\beta \beta} S D^{\beta \beta} +   D^{\beta \alpha} S D^{\alpha \beta} = D^{\beta \beta} 
\end{equation}

After that, we have:
\begin{equation}
    \mathrm{Tr}(D^{\alpha \alpha}S D^{\alpha \alpha}S) +  \mathrm{Tr}(D^{\beta \beta}S D^{\beta \beta}S) =N -2 \mathrm{Tr}(D^{\alpha \beta} S D^{\beta \alpha}S)
\end{equation}
By substitution, we have:
\begin{equation}
      \sum_{ij} W_{ijji} = \frac{1}{2}N -2 \mathrm{Tr}(D^{\alpha \beta} S D^{\beta \alpha}S)+2\mathrm{Tr}(D^{\alpha \alpha}S D^{\beta \beta} S)
\end{equation}
Then, we have:
\begin{equation}
    \langle \hat{S}^2 \rangle_0 = \frac{1}{2}N +\frac{1}{4}[\mathrm{Tr}(S(D^{\alpha \alpha}-D^{\beta \beta}))]^2 + \mathrm{Tr}(S D^{\alpha \beta})\times \mathrm{Tr}(S D^{\beta \alpha})+ \mathrm{Tr}(D^{\alpha \beta} S D^{\beta \alpha}S)-\mathrm{Tr}(D^{\alpha \alpha}S D^{\beta \beta} S)
\end{equation}

\subsection*{Appendix B: Derivation from Eq. \ref{eq:s1def} to Eq. \ref{eq:s1}}

By the definition of $W$ in eq. \ref{eq:wdef}, we have:
\begin{equation}
    \begin{split}
        \langle X | \hat{S}^2 |Y \rangle &= [\frac{1}{2}\left(S_{ai}-S_{\bar{a}\bar{i}}\right) \left( \sum_j \left( S_{jj}-S_{\bar{j}\bar{j}}  \right)\right) + S_{a\bar{i}} (\sum_j S_{\bar{j}j}) +S_{\bar{a}i} (\sum_j S_{j\bar{j}})\\
        &-\frac{1}{2} \sum_j \left(  \left( S_{aj}-S_{\bar{a}\bar{j}}  \right)  \left( S_{ji}-S_{\bar{j}\bar{i}}  \right)  \right) - \sum_j S_{a \bar{j}} S_{\bar{j}i} - \sum_j S_{\bar{a}j} S_{j \bar{i}}]^*
    \end{split}
\end{equation}
We derive the most complicated term as an example, and other terms can be derived using similar ideas.
\begin{equation}
    \begin{split}
        \sum_j \left(  \left( S_{aj}-S_{\bar{a}\bar{j}}  \right)  \left( S_{ji}-S_{\bar{j}\bar{i}}  \right)  \right) &= \sum_j \left( \left( \left(C^{\alpha}_a\right)^\dagger
  S C_j^\alpha- \left(C^{\beta}_a \right)^\dagger S C_j^\beta      \right)   \right)
  \left( \left(C^{\alpha}_j\right)^\dagger
  S C_i^\alpha- \left(C^{\beta}_j \right)^\dagger S C_i^\beta      \right) \\
    \end{split}
\end{equation}
We treat this term as follows:
\begin{equation}
    \sum_j (C_a^\sigma)^\dagger S C_j^\sigma (C_j^{\sigma^\prime})^\dagger S C_i^{\sigma^\prime} = (C_a^\sigma)^\dagger \sum_j (S C_j^\sigma (C_j^{\sigma^\prime})^\dagger S) C_i^{\sigma^\prime} = (C_a^\sigma)^\dagger S (D^{\sigma^\prime \sigma})^T S C_i^{\sigma^\prime}
\end{equation}

\subsection*{Appendix C: Proof of Sharing Eigenvectors between Eq. \ref{eq:eoms} and Eq. \ref{eq:eomw}}

The two equations eq. \ref{eq:eoms} and eq. \ref{eq:eomw} are both matrix representations of the Liouvillian in a projected subspace. We define the Liouvillian as:
\begin{equation}
    \mathcal{L}_K[\cdot]=[K, \cdot]
\end{equation}
Because the Hamiltonian is free of spin, we have:
\begin{equation}
    [\mathcal{L}_{\hat{H}},\mathcal{L}_{\hat{S}^2}] (\cdot) = \mathcal{L}_{\hat{H}}(\mathcal{L}_{\hat{S}^2}(\cdot)) -  \mathcal{L}_{\hat{S}^2}(\mathcal{L}_{\hat{H}}(\cdot))=[\hat{H},[\hat{S}^2,\cdot]]-[\hat{S}^2,[\hat{H},\cdot]]=[[\hat{H},\hat{S}^2],\cdot]= \mathcal{L}_{[\hat{H},\hat{S}^2]}(\cdot)=0
\end{equation}
Only when the operator basis is complete will the corresponding matrix representation share the same eigenvectors.

\subsection*{Appendix D: Geometry of Calculated Molecules}

\paragraph{H$_2$O (\(N=3\)).}
Bond length \(r=0.9572\,\mathrm{\AA}\), bond angle \(104.5^{\circ}\); O at the origin.
\begin{center}
\begin{tabular}{l r r r}
\hline
Atom & $x$ & $y$ & $z$\\
\hline
O & 0.0000000000 & 0.0000000000 & 0.0000000000\\
H & 0.0000000000 & 0.0000000000 & 0.9572000000\\
H & 0.9267109214 & 0.0000000000 & -0.2396637399\\
\hline
\end{tabular}
\end{center}

\paragraph{H$_2$O$^+$ (\(N=3\)).}
Same geometry as H$_2$O; charge = +1, spin = 1 (doublet).
\begin{center}
\begin{tabular}{l r r r}
\hline
Atom & $x$ & $y$ & $z$\\
\hline
O & 0.0000000000 & 0.0000000000 & 0.0000000000\\
H & 0.0000000000 & 0.0000000000 & 0.9572000000\\
H & 0.9267109214 & 0.0000000000 & -0.2396637399\\
\hline
\end{tabular}
\end{center}

\paragraph{H$_3$ (\(N=3\)).}
Equilateral triangle in the $xy$-plane; side length \(\ell=0.9088\,\mathrm{\AA}\).
\begin{center}
\begin{tabular}{l r r r}
\hline
Atom & $x$ & $y$ & $z$\\
\hline
H & 0.0000000000 & 0.5246959246 & 0.0000000000\\
H & -0.4544000000 & -0.2623479623 & 0.0000000000\\
H & 0.4544000000 & -0.2623479623 & 0.0000000000\\
\hline
\end{tabular}
\end{center}

\end{document}